\documentclass[12pt,draftcls,onecolumn]{IEEEtran}
\usepackage{amsfonts}
\usepackage[tbtags]{amsmath}
\usepackage{graphicx,subfigure}
\usepackage{multirow}
\usepackage{color}      
\usepackage{epsfig}
\usepackage{amssymb}
\usepackage{tipa}
\usepackage{bbm}
\def\endproof{\hspace*{\fill}~$\blacksquare$}
\long\def\comment#1{}

\newcommand{\beq}{\begin{equation}}
\newcommand{\eeq}{\end{equation}}
\newcommand{\beqno}{\begin{equation*}}
\newcommand{\eeqno}{\end{equation*}}

\newcommand{\bes}{\begin{split}}
\newcommand{\ees}{\end{split}}
\newcommand{\bdm}{\begin{displaymath}}
\newcommand{\edm}{\end{displaymath}}

\newcommand{\goes}{\rightarrow}

\newtheorem{theorem}{Theorem}

\newtheorem{lemma}{Lemma}

\newtheorem{definition}{Definition}

\newcommand{\bd}{\begin{definition}}
\newcommand{\ed}{\end{definition}}

\newcommand{\bv}{\begin{vugraph}}
\newcommand{\ev}{\end{vugraph}}
\newcommand{\bi}{\begin{itemize}}
\newcommand{\ei}{\end{itemize}}
\newcommand{\ben}{\begin{enumerate}}
\newcommand{\een}{\end{enumerate}}

\newcommand{\bean}{\begin{eqnarray*} }
\newcommand{\eean}{\end{eqnarray*} }
\newcommand{\bea}{\begin{eqnarray} }
\newcommand{\eea}{\end{eqnarray} }

\newcommand{\ba}{\begin{array} }
\newcommand{\ea}{\end{array} }

\begin{document}

\title{Rate and power allocation under the pairwise distributed source coding
constraint}

\author{\authorblockN{Shizheng Li and Aditya Ramamoorthy}\\
\authorblockA{Department of Electrical and Computer Engineering\\
Iowa State University\\
Ames, Iowa 50011\\
Email: \{szli, adityar\}@iastate.edu}
\thanks{The material in this work was presented in part at the IEEE Intl. Symp. on Info. Th. 2008.
} \thanks{This research was supported in part by NSF grant
CNS-0721453. }}

\maketitle
\begin{abstract}
We consider the problem of rate and power allocation for a sensor
network under the pairwise distributed source coding constraint.
For noiseless source-terminal channels, we show that the minimum
sum rate assignment can be found by finding a minimum weight
arborescence in an appropriately defined directed graph. For
orthogonal noisy
 source-terminal channels, the minimum sum power allocation can be found by finding
 a minimum weight matching forest in a mixed graph. Numerical results are presented for both
cases showing that our solutions always outperform previously
proposed solutions. The gains are considerable when source
correlations are high.
\end{abstract}


\section{Introduction}
\label{sec:intro} The availability of low-cost sensors has enabled
the emergence of large-scale sensor networks in recent years.
Sensor networks typically consist of sensors that have limited
power and are moreover energy constrained since they are usually
battery-operated. The data that is sensed by sensor networks and
communicated to a terminal\footnote{We shall use terminal and sink
interchangably throughout this paper.} is usually correlated.
Thus, for sensor networks it is important to allocate resources
such as rates and power by taking the correlation into account.
The famous Slepian-Wolf theorem \cite{slepianwolf} shows that the
distributed compression (or distributed source coding) of
correlated sources can in fact be as
efficient as joint compression. 
Coding techniques that approach the Slepian-Wolf bounds have been
investigated \cite{pradhandiscus} and their usage proposed in
sensor networks
\cite{xiongspmag}. 
Typically one wants to minimize metrics such as the total rate or
total power expended by the sensors in such situations. A number
of authors have considered problems of this flavor
\cite{razvan,ramamoorthy07,cristescuBV05}. These papers assume the
existence of Slepian-Wolf codes that work for a large number of
sensors.

In practice, the design of low-complexity Slepian-Wolf codes is
well understood only for the case of two sources (denoted $X$ and
$Y$) and there have been constructions that are able to operate on
the boundary of the Slepian-Wolf region. In particular, the design
of codes
(eg.\cite{liverisXG02},\cite{aaronG02},\cite{schonbergRP02}) is
easiest for the corner points (asymmetric Slepian-Wolf coding)
where the rate pair is either $(H(X), H(Y|X))$ or $(H(X|Y),
H(Y))$. Several symmetric code designs are proposed in
\cite{schonbergRP04},\cite{totozarasoaRG08},\cite{baiYBH08} in
which the authors mainly focus on two correlated sources.
In \cite{liverisXG02}, the correlation between two binary sources
are assumed to be symmetric and the LDPC code is designed for a
virtual BSC correlation channel, while the codes designed in
\cite{schonbergRP02}, \cite{schonbergRP04} and
\cite{totozarasoaRG08} are suitable for arbitrary correlation
between the two binary sources. The authors of
\cite{stankovicLXG06} proposed code designs for multiple sources.
For two uniformly distributed binary sources whose correlation can
be modeled as a BSC channel, their design supports both symmetric
and asymmetric coding and approaches Slepian-Wolf bound. However,
when it comes to more than two sources, in order to achieve
optimum rate (joint entropy), they have a strong assumption on
correlation model, i.e., the correlation between all the sources
is solely described by their modulo-2 sum. Thus, given the current
state of the art in code design it is of interest to consider
coding strategies for sensor networks where pairs of nodes can be
decoded at a time instead of all at once. This observation was
made in the work of Roumy and Gesbert in \cite{roumyG07jour}. In
that work they formulated the pairwise distributed source coding
problem and presented algorithms for rate and power allocation
under different scenarios. In particular, they considered the case
when there exist direct channels between each source node and the
terminal. Furthermore, the terminal can only decode the sources
pairwise. We briefly review their work below. The work of
\cite{roumyG07jour} considers two cases.
\begin{itemize}
\item[i)] {\it Case 1 - Noiseless node-terminal channels.}\\
Under this scenario, they considered the problem of deciding which
particular nodes should be decoded together at the terminal and
their corresponding rate allocations so that the total sum rate is
minimized. \item[ii)] {\it Case 2 - Orthogonal noisy node-terminal
channels.}\\
In this case the channels were assumed to be noisy and orthogonal
and the objective was to decide which nodes would be paired so
that overall power consumption is minimized.
\end{itemize}
In \cite{roumyG07jour}, the problem was mapped onto the problem of
choosing the minimum weight matching \cite{kleinbergT05} of an
appropriately defined weighted undirected graph. Each node
participate in joint decoding only once.

In this paper we consider a class of pairwise distributed source
coding solutions that is larger than the ones considered in
\cite{roumyG07jour}. The basic idea is that previously decoded
data can be used as side information for other sources. A simple
example demonstrates that it is not necessary to only consider
matchings
Consider four correlated sources $X_1, X_2, X_3$ and $X_4$. The
solution of \cite{roumyG07jour} constructs a complete graph on the
four nodes $X_1, \dots, X_4$ and assigns the edge weights as the
joint entropies i.e. the edge $(X_i, X_j)$ is assigned weight
$H(X_i, X_j)$. A minimum weight matching algorithm is then run on
this graph to find the minimum sum rate and the rate allocation.
Suppose that this yields the matching $(X_1, X_3)$ and $(X_2,
X_4)$ so that the sum rate becomes \beqno \sum_{i = 1}^4 R_i =
H(X_1, X_3) + H(X_2, X_4). \eeqno Since conditioning reduces
entropy, it is simple to observe that \beqno
\begin{split}
H(X_1, X_3) + H(X_2, X_4) \geq H(X_1) + H(X_3|X_1) + H(X_2|X_3) +
H(X_4 |X_2).
\end{split}
\eeqno We now show that an alternative rate allocation: $R_1 =
H(X_1), R_2 = H(X_2|X_3), R_3 = H(X_3|X_1)$ and $R_4 = H(X_4|X_2)$
can still allow pairwise decoding of the sources at the terminal.
Note that at the decoder we have,
\begin{itemize}
\item[a)] $X_1$ is known since $R_1 = H(X_1)$. \item[b)]$X_3$ can
be recovered by jointly decoding for $X_3$ and $X_1$ since $X_1$
is known and the decoder has access to $H(X_3|X_1)$ amount of
data. \item[c)] $X_2$ can be recovered since $X_3$ is known (from
above) and the decoder has access to $H(X_2|X_3)$ amount of data.
\item[d)] Similarly, $X_4$ can be recovered.
\end{itemize}
As we see above, the sources can be decoded at the terminal in a
pipelined manner. Note that we can leverage the coding solutions
proposed for two sources at the corner points in this case since
the encoder for $X_3$ can be designed assuming that $X_1$ is known
perfectly, the encoder for $X_2$ can be designed assuming that
$X_3$ is known perfectly etc. The method of source-splitting
\cite{rimoldi_Source_Sp, source_splitting} is closely related to
this approach. Given $M$ sources and an arbitrary rate point in
their Slepian-Wolf region, it converts the problem into a rate
allocation at a Slepian-Wolf corner point for appropriately
defined $2M - 1$ sources. However as pointed out before, code
designs even for corner points are not that well understood for
more than two sources. Thus, while using source-splitting can
result in sum-rate optimality i.e. the sum rate is the joint
entropy, it may not be very practical given the current state of
the art. Moreover, for $M$ sources it requires the design of
approximately twice as many encoders and more decoding sub-modules
that also comes at the cost of complexity.


In this paper, motivated by complexity issues, we present an
alternate formulation of the pairwise distributed source coding
problem that is more general than \cite{roumyG07jour}. We
demonstrate that for noiseless channels the minimum sum rate
allocation problem becomes one of finding a minimum weight
arborescence of an appropriately defined directed graph. Next, we
show that in the case of noisy channels, the minimum sum power
allocation problem can be mapped onto finding the minimum weight
matching forest of an appropriately defined mixed graph\footnote{A
mixed graph has both directed and undirected edges}. Simulation
results show that our solutions are significantly better than
those in \cite{roumyG07jour} in the cases when correlations are
high.

This paper is organized as follows. We formulate the problem and
briefly review previous solutions based on matching in Section
\ref{sec:formulation}. In Section \ref{sec:noiseless-case} and
\ref{sec:noisy-case} we present our solution for noiseless
channels and noisy channels respectively. Numerical results for
the both cases are given in Section \ref{sec:results} and Section
\ref{sec:conclusion} concludes this paper.

\section{Problem formulation and overview of related work}
\label{sec:formulation}
\par Consider a set of correlated sources $X_1, X_2, \dots, X_n$ transmitting data to one sink in a wireless sensor network. We assume that every source can transmit data directly to the terminal. 
The source $X_i$ compresses its data at rate $R_i$ and sends it to the sink. We assume that the sources encode only their own data. Furthermore, we consider the class of solutions where the sink can recover a given source with the help of at most one other source. 
The problem has two cases.
\begin{itemize}
\item[i)] {\it Case 1 - Noiseless node-terminal channels.}\\
Assume that there is no noise in the channel. In order to reduce
the storage requirement at the sensors, we want to minimize the
sum rate, i.e., $\min \sum_{i=1}^{n} R_i$. \item[ii)] {\it Case 2
- Orthogonal noisy node-terminal
channels.}\\
Assume that channels between sources and sink are corrupted by
additive white Gaussian noise and there is no internode
interference. In this case, source channel separation holds
\cite{barrosS06}. The capacity of the channel between node $i$ and
the sink with transmission power $P_i$ and channel gain $\gamma_i$
is
$C_i(P_i)\triangleq \log(1+\gamma_iP_i)$,
where noise power is normalized to one and channel gains are constants known to the terminal. Rate $R_i$ should satisfy $R_i\leq C_i(P_i)$. Let $[n]$ denote the index set $\{1,\ldots,n\}$. The transmission power is constrained by peak power constraint:$\forall i\in [n], P_i\leq P_{max}$. 
In this context, our objective is to minimize the sum power ,
i.e., $\min \sum_{i=1}^{n} P_i$. Note that in the implementation
from the practical point of view, we can use joint distributed
source coding and channel coding \cite{GFZZ07,ZhongGF05}, once the
pairing of nodes involved in jointly decoding are known from the
resource allocation solution.

\end{itemize}
\par We now overview the work of \cite{roumyG07jour}.
For noiseless case, in order for the terminal to recover data
perfectly, the rates for a pair of nodes $i$ and $j$ should be in
the Slepian and Wolf region
\begin{equation*}
SW_{ij}\triangleq \left\{ (R_i,R_j) : R_i\geq H(X_i|X_j), R_j\geq
H(X_j|X_i),R_i+R_j\geq H(X_i,X_j) \right\}.
\end{equation*}
Note that $H(X_i,X_j)$ is the minimum sum rate while $i$ and $j$
are paired to perform joint decoding. The matching solution of the
problem is as follows. Construct an undirected complete graph $G=
(V,E)$ , where $|V| = n$. Let $W_E(i,j)$ denote weight on
undirected edge $(i,j)$, $W_E(i,j)=H(X_i,X_j)$. Then, find a
minimum weight matching $\mathcal{P}$ of $G$. For $(i,j)\in
\mathcal{P}$, the optimal rate allocation $(R_i, R_j)$ can be any
point on the slope of the SW region of nodes $i$ and $j$ since
they give same sum rate for a pair. We can simply set $(R_i, R_j)$
for $(i,j)\in \mathcal{P}$ to be either $(H(X_i), H(X_j|X_i))$ or
$(H(X_j), H(X_i|X_j))$, i.e., at the corner points of SW region.

For noisy case, the rate region for a pair of nodes is the
intersection of SW region and capacity region $C_{ij}$:
$C_{ij}(P_i,P_j)\triangleq \{(R_i, R_j): R_i\leq C_i(P_i), R_j\leq
C_j(P_j)\}$.
It is easy to see that for a node $i$ with rate $R_i$ and power
$P_i$, at the optimum $R_i^*=C_i(P_i^*)$, i.e. the inequality
$R_i\leq C_i(P_i)$ constraint is met with equality. Thus, the
power assignment is given by the inverse function of $C_i$ which
we denote by $Q_i(R_i)$, i.e.,  $P_i^*=
Q_i(R_i^*)=(2^{R_i^*}-1)/\gamma_i$. This problem can also be
solved by finding minimum matching on a undirected graph. However
the weights in this case are the minimum sum power for each pair
of nodes. The solution has two steps:
\begin{enumerate}
\item Find optimal rate-power allocations for all possible node
pairs: $\forall (i,j)\in [n]^2$ s.t. $i<j$:
\begin{equation}\label{eq:NoisyRGS1}
(R_{ij}^*(i),R_{ij}^*(j))=\arg\min Q_i(R_{ij}(i))+Q_j(R_{ij}(j))
\end{equation}
\begin{equation}
s.t. (R_{ij}(i),R_{ij}(j))\in SW_{ij}\cap C_{ij}(P_{max},P_{max})
\end{equation}
The power allocations are given by $P_{ij}^*(i)=Q_i(R_{ij}^*(i))$
and $P_{ij}^*(j)=Q_j(R_{ij}^*(j))$. The rates
$R_{ij}(i),R_{ij}(j)$ are the rates for node $i$ and node $j$ when
$i$ and $j$ are paired. Note that when $i$ and another node $k\neq
j$ are considered as a pair, the rate for $i$ may be
different,i.e., $R_{ij}(i)\neq R_{ik}(i)$. \item Construct an
undirected complete graph $G= (V,E)$, where
$W_E(i,j)=P_{ij}^*(i)+P_{ij}^*(j)$ for edge $(i,j)$, and find a
minimum matching $\mathcal{P}$ in $G$. The power allocation for
node pair $(i,j)\in \mathcal{P}$ denoted by $(P_i, P_j)$ is
$(P_{ij}^*(i),P_{ij}^*(j))$ and the corresponding rate allocation
can be found.
\end{enumerate}
The solution for step (1) is given in \cite{roumyG07jour} and
denoted as $(P_{ij}^*(i), P_{ij}^*(j), R_{ij}^*(i), R_{ij}^*(j))$.
This solution is the optimum rate-power allocation between a pair
of nodes $i$ and $j$ under the peak power constraint and SW region
constraint. Note that in this case, the rate assignments for $i$
and $j$ do not
necessarily happen at the corner of the SW region.

\section{Noiseless case}
\label{sec:noiseless-case}
As shown by the example in Section \ref{sec:intro}, the rate allocation given by matching may not be optimum and in fact there exist other schemes that have a lower rate while still working with the current coding solutions to the two source SW problem. 
We now present a formal definition of the pairwise decoding
constraint.

\begin{definition} \label{def:pairwise_prop} {\it Pairwise property of rate assignment.}
Consider a set of discrete memoryless sources $X_1, X_2, \dots,
X_n$ and the corresponding rate assignment $\mathbf{R} = (R_1,
R_2, \dots, R_n)$. The rate assignment is said to satisfy the
pairwise property if for each source $X_i, i\in [n]$, there exists
an ordered sequence of sources $(X_{i_1}, X_{i_2}, \dots,
X_{i_k})$ such that
\begin{align}
R_{i_1} &\geq H(X_{i_1}), \label{eq:pairwise_prop_1}\\
R_{i_j} & \geq H(X_{i_j} | X_{i_{j-1}}) \text{,~~ for $2 \leq j
\leq k$, and }  \label{eq:pairwise_prop_2}\\
R_{i} & \geq H(X_{i} | X_{i_{k}}).
\label{eq:pairwise_prop_3}\vspace{-2mm}
\end{align}

\end{definition}
Note that a rate assignment that satisfies the pairwise property
allows the possibility that each source can be reconstructed at
the decoder by solving a sequence of decoding operations at the SW
corner points e.g. for decoding source $X_i$ one can use $X_{i_1}$
(since $R_{i_1} \geq H(X_{i_1})$), then decode $X_{i_2}$ using the
knowledge of $X_{i_1}$. Continuing in this manner finally $X_i$
can be decoded. A rate assignment $\mathbf{R}$ shall be called
pairwise valid (or valid in this section), if it satisfies the
pairwise property. In this section, we focus on looking for a
valid rate allocation that minimizes the sum rate. An equivalent
definition can be given in graph-theoretic terms by constructing a
graph called the pairwise property test graph corresponding to the
rate assignment.

\noindent {\it Pairwise Property Test Graph Construction}
\begin{enumerate}
\item Inputs : the number of nodes $n$, $H(X_i)$ for all
$i\in[n]$, $H(X_i | X_j)$ for all $i,j\in [n]^2$ and the rate
assignment $\mathbf{R}$. \item Initialize a graph $G = (V, A)$
with a total of $2n$ nodes i.e. $|V| = 2n$. There are $n$ {\it
regular} nodes denoted $1, 2, \dots, n$ and $n$ {\it starred}
nodes denoted $1^*, 2^*, \dots , n^*$. \item Let $W_A(j\rightarrow
i)$ denote the weight on directed edge $(j\rightarrow i)$. For
each $i \in [n]$:
\begin{itemize}
\item[i)] If $R_i \geq H(X_i)$ then insert edge $(i^* \goes i)$
with $W_A(i^* \goes i)=H(X_i)$. \item[ii)] If $R_i \geq H(X_i |
X_j)$ then insert edge $(j \goes i)$ with $W_A(j \goes i)=H(X_i |
X_j)$.
\end{itemize}
\item Remove all nodes that do not participate in any edge.
\end{enumerate}

We denote the resulting graph for a given rate allocation by
$G(\mathbf{R}) = (V, A)$. Note that if $\mathbf{R}$ is valid, the
graph still contains at least one starred node. Next, based on
$G(\mathbf{R})$ we define a set of nodes that are called the
parent nodes. $\text{Parent}(\mathbf{R}) = \{i^* | (i^* \goes i)
\in A\}$, i.e., $\text{Parent}(\mathbf{R})$ corresponds to the
starred nodes for the set of sources for which the rate allocation
is at least the entropy. Mathematically if $i^* \in
\text{Parent}(\mathbf{R})$, then $R_{i} \geq H(X_i)$. We now
demonstrate the equivalence between the pairwise property and the
construction of the graph above.
\begin{lemma}\label{lem:Noiseless PP2path}
Consider a set of discrete correlated sources $X_1, \dots X_n$ and
a corresponding rate assignment $\mathbf{R} = (R_1, \dots, R_n)$.
Construct $G(\mathbf{R})$ based on the algorithm above. The rate
assignment $\mathbf{R}$ satisfies the pairwise property if and
only if for all regular nodes $i \in V$ there exists a starred
node $j^* \in \text{Parent}(\mathbf{R})$ such that there exists
directed path from $j^*$ to $i$ in $G(\mathbf{R})$.
\end{lemma}
\emph{Proof:} Suppose that $G(\mathbf{R})$ is such that for all
regular nodes $i \in V$, there exists a  $j^* \in
\text{Parent}(\mathbf{R})$ so that there is a directed path from
$j^*$ to $i$. We show that this implies the pairwise property for
$X_i$. Let the path from $j^*$ to $i$ be denoted $j^* \goes j
\goes \alpha_1 \dots \goes \alpha_k \goes i$. We note that $R_j
\geq H(X_j)$ by construction. Similarly edge $(\alpha_l \goes
\alpha_{l+1})$ exists in $G(\mathbf{R})$ only because
$R_{\alpha_{l+1}} \geq H(X_{\alpha_{l+1}} | X_{\alpha_{l}})$ and
likewise $R_i \geq H(X_{i} | X_{\alpha_{k}})$. Thus for source $i$
we have found the ordered sequence of sources $(X_j, X_{\alpha_1},
\dots, X_{\alpha_k})$ that satisfy properties
(\ref{eq:pairwise_prop_1}), (\ref{eq:pairwise_prop_2}) and
(\ref{eq:pairwise_prop_3}) in definition \ref{def:pairwise_prop}.

Conversely, if $\mathbf{R}$ satisfies the pairwise property, then
for each $X_i$, there exists an ordered sequence $(X_{i_1}, \dots,
X_{i_k})$ that satisfies properties (\ref{eq:pairwise_prop_1}),
(\ref{eq:pairwise_prop_2}) and (\ref{eq:pairwise_prop_3}) from
definition \ref{def:pairwise_prop}. This implies that there exists
a directed path from $i_1^*$ to $i$ in $G(\mathbf{R})$, since
$(i_1^* \goes i_1) \in A$ because $R_{i_1} \geq H(X_{i_1})$ and
furthermore $(i_{j-1} \goes i_j) \in A$ because $R_{i_j} \geq
H(X_{i_j} | X_{i_{j-1}})$, for $j=2, \ldots, k$.
\endproof

We define another set of graphs that are useful for presenting the
main result of this section.
\begin{definition} {\it Specification of $G_{i^*}(\mathbf{R})$.}
Suppose that we construct graph $G(\mathbf{R})$ as above and find
$\text{Parent}(\mathbf{R})$. For each $i^* \in
\text{Parent}(\mathbf{R})$ we construct $G_{i^*}(\mathbf{R})$ in
the following manner: For each $j^* \in \text{Parent}(\mathbf{R})
\backslash \{i^*\}$ remove the edge $(j^* \goes j)$ and the node
$j^*$ from $G(\mathbf{R})$.
\end{definition}

For the next result we need to introduce the concept of an
arborescence \cite{kleinbergT05}.

\begin{definition}
An \textit{arborescence (also called directed spanning tree)} of a
directed graph $G=(V,A)$ rooted at vertex $r \in V$ is a subgraph
$T$ of $G$ such that it is a spanning tree if the orientation of
the edges is ignored and there is a path from $r$ to all $v \in V$
when the direction of edges is taken into account.
\end{definition}

\begin{theorem}\label{theo:noiseless1}
Consider a set of discrete correlated sources $X_1, \dots, X_n$
and let the corresponding rate assignment $\mathbf{R}$ be pairwise
valid. Let $G(\mathbf{R})$ be constructed as above. There exists
another valid rate assignment $\mathbf{R}^{'}$ that can be
described by the edge weights of an arborescence of
$G_{i^*}(\mathbf{R})$ rooted at $i^*$ where $i^* \in
\text{Parent}(\mathbf{R})$ such that $R^{'}_j \leq R_j$, for all
$j \in [n]$.
\end{theorem}
\emph{Proof:} We shall show that a new subgraph can be constructed
from which $\mathbf{R}^{'}$ can be obtained. This shall be done by
a series of graph-theoretic transformations.

Pick an arbitrary starred node $j^* \in \text{Parent}(\mathbf{R})$
and construct $G_{j^*}(\mathbf{R})$. We claim that in the current
graph $G_{j^*}(\mathbf{R})$ there exists a path from the starred
node $j^*$ to all regular nodes $i \in [n]$. To see this note that
since $\mathbf{R}$ is pairwise valid, for each regular node $i$
there exists a path from some starred node to $i$ in
$G(\mathbf{R})$. If for some regular node $i$, the starred node is
$j^*$, the path is still in $G_{j^*}(\mathbf{R})$. Now consider a
regular node $i_1$ and suppose there exists a directed path $k^{*}
\goes k \goes \beta_1 \dots \goes i_1$ in $G(\mathbf{R})$ where
$k^* \in \text{Parent}(\mathbf{R}), k^* \neq j^*$. Since $k^* \in
\text{Parent}(\mathbf{R})$, $R_k \geq H(X_k) \geq H(X_k | X_l)
\text{~~ } \forall l \in [n]$. This implies that edge $(l \goes
k)$ is in $G_{j^*}(\mathbf{R}),\forall l \in [n]$, in particular,
$(j \goes k)\in G_{j^*}(\mathbf{R})$. Therefore, in
$G_{j^*}(\mathbf{R})$ there exists the path $j^* \goes j \goes
k\goes \beta_1 \dots \goes i_1$. This claim implies that there
exists an arborescence rooted at $j^*$ in $G_{j^*}(\mathbf{R})$
\cite{kleinbergT05}.

Suppose we find such one such arborescence $T_{j^*}$ of
$G_{j^*}(\mathbf{R})$. In $T_{j^*}$ every node except $j^*$ has
exactly one incoming edge (by the property of an arborescence
\cite{kleinbergT05}). Let $inc(i)$ denote the node such that
$(inc(i) \goes i) \in T_{j^*}$. We define a new rate assignment
$\mathbf{R}^{'}$ as
\begin{align*} R_{i}^{'} &= W_A(inc(i) \goes i) = H(X_i | X_{inc(i)}) \text{~~(for $i \in [n]$ and $i\neq j$), and}\\
R_{j}^{'} &=W_A(j^*\goes j)= H(X_j).
\end{align*}
The existence of edge $(j^*\goes j)\in G(\mathbf{R})$ implies
$R_{j}^{'} = H(X_j)\leq R_j$. Similarly, we have $R_i^{'}\leq R_i$
for $i \in [n]\backslash \{j\}$. And it is easy to see that
$\mathbf{R}^{'}$ is a valid rate assignment.
\endproof

Thus, the above theorem implies that valid rate assignments that
are described on arborescences of the graphs $G_{i^*}(\mathbf{R})$
are the best from the point of view of minimizing the sum rate.
Finally we have the following theorem that says that the valid
rate assignment that minimizes the sum rate can be found by
finding minimum weight arborescences of appropriately defined
graphs. For the statement of the theorem we need to define the
following graphs.
\begin{itemize}
\item[a)] The graph $G^{tot} = (V^{tot}, A^{tot})$ is such that
$V^{tot}$ consists of $n$ regular nodes $1, \dots, n$ and $n$
starred nodes $1^*, \dots , n^*$, $|V^{tot}| = 2n$. The edge set
$A^{tot}$ consists of edges $(i^* \goes i), W_A(i^* \goes i) =
H(X_i)$ for $i\in[n]$ and edges $(i \goes j), W_A(i \goes j) =
H(X_j | X_i)$ for all $i, j\in[n]^2$. \item[b)] For each $i = 1,
\dots, n$ we define $G_{i^*}$ as the graph obtained from $G^{tot}$
by deleting all edges of the form $(j^* \goes j)$ for $j \neq i$
and all nodes in $\{1^*, \dots, n^*\} \backslash \{i^*\}$.
\end{itemize}

\begin{theorem}\label{theo:noiseless 2}
Consider a set of sources $X_1, \dots, X_n$. Suppose that we are
interested in finding a valid rate assignment $\mathbf{R} = (R_1,
\dots, R_n)$ for these sources so that the sum rate $\sum_{i=1}^n
R_i$ is minimum. Let $\mathbf{R}^{i^*}$ denote the rate assignment
specified by the minimum weight arborescence of $G_{i^*}$. Then
the optimal valid rate assignment can be found as \vspace{-2mm}
\beqno R_{opt} = \arg \min_{i \in \{1, \dots, n\}} \sum_{j=1}^n
R_{j}^{i^*} \eeqno
\end{theorem}
\emph{Proof.} From Theorem \ref{theo:noiseless1} we have that any
valid rate assignment $\mathbf{R}$ can be transformed into new
rate assignment that can be described on an arborescence of
$G_{i^*}(\mathbf{R})$ rooted at $i^*$ and suitable weight
assignment. It is component-wise lower than $\mathbf{R}$. This
implies that if we are interested in a minimum sum rate solution,
it suffices to focus our attention on solutions specified by all
solutions that can be described by all possible arborescences of
graphs of the form $G_{i^*}(\mathbf{R})$ over all $i^* = 1^*,
\dots, n^*$ and all possible valid rate assignments $\mathbf{R}$.
\par Now consider the graph $G_{i^*}$ defined above. We note that
all graphs of the form $G_{i^*}(\mathbf{R})$ where $\mathbf{R}$ is
valid are subgraphs of $G_{i^*}$. Therefore finding the minimum
cost arborescence of $G_{i^*}$ will yield us the best rate
assignment possible within the class of solutions specified by
$G_{i^*}(\mathbf{R})$. Next, we find the best solutions
$\mathbf{R}^{i^*}$ for all $i \in [n]$ and pick the solution with
the minimum cost. This yields the optimal rate assignment.
\endproof

\section{Noisy case}
\label{sec:noisy-case} In this section we consider the case when
the sources are connected to the terminal by orthogonal noisy
channels. In this case, the objective is to minimize the sum
power. Therefore the optimum rate allocation within a pair of
sources may not be at the corner points of SW region. We want some
node pairs working at corner points while some others working on
the slope of the SW region. Taking this into account, we
generalize the concept of pairwise property.

For a given rate assignment $\mathbf{R}$, we say that $X_i$ is
\textit{initially decodable} if $R_i \geq H(X_i)$, or together
with another source $X_j$, $(R_i, R_j)\in SW_{ij}$. If $R_i \geq
H(X_i)$, it can be decoded by itself. If $(R_i, R_j)\in SW_{ij}$,
SW codes can be designed for $X_i,X_j$ and they can be recovered
by joint decoding. In addition, if we take advantage of previously
decoded source data to help decode other sources as we did in the
noiseless case, starting with an initially decodable source, more
sources can potentially be recovered.

\begin{definition} \label{def:ged pairwise_prop} {\it Generalized pairwise property of rate assignment.}
Consider a set of discrete memoryless sources $X_1, \dots, X_n$
and the corresponding rate assignment $\mathbf{R} = (R_1, \dots,
R_n)$. The rate assignment is said to satisfy the generalized
pairwise property if for each $X_i, i\in [n]$, $X_i$ is initially
decodable, or there exists an ordered sequence of sources
$(X_{i_1}, X_{i_2}, \dots, X_{i_k})$ such that \vspace{-2mm}
\begin{align}
X_{i_1}&   \text{is initially decodable}, \label{eq:ged pairwise_prop_1}\\
R_{i_j}&  \geq H(X_{i_j} | X_{i_{j-1}}), \text{~~ for $2 \leq j
\leq k$.}  \label{eq:ged pairwise_prop_2}\\
R_{i}&  \geq H(X_{i} | X_{i_{k}})  \label{eq:ged pairwise_prop_3}
\end{align}
\end{definition}
A rate assignment $\mathbf{R}$ shall be called generalized
pairwise valid (or valid in this section), if it satisfies the
generalized pairwise property and for every rate $R_i\in
\mathbf{R}$, $Q_i(R_i)\leq P_{max}$. A valid rate assignment
allows every source to be recovered at the sink. A power
assignment $\mathbf{P}=(P_1, P_2, \dots, P_n)$ shall be called
valid, if the corresponding rate assignment is valid.

We shall introduce generalized pairwise property test graph. The
input and initialization are the same as pairwise property test
graph construction. Then, for each $i \in [n]$:
\begin{itemize}
\item[i)] If $R_i \geq H(X_i)$ then insert directed edge $(i^*
\goes i)$ with weight $W_A(i^* \goes i)=Q_i(H(X_i))$. \item[ii)]
If $R_i \geq H(X_i | X_j)$ then insert directed edge $(j \goes i)$
with weight $W_A(j \goes i)=Q_i(H(X_i|X_j))$. \item[iii)] If
$(R_i, R_j)\in SW_{ij}$, then insert undirected edge $(i,j)$ with
weight
$W_E(i,j)=Q_i(R_{ij}^*(i))+Q_j(R_{ij}^*(j))=P_{ij}^*(i)+P_{ij}^*(j)$.
Note that as pointed out in Section \ref{sec:formulation},
$(P_{ij}^*(i), P_{ij}^*(j), R_{ij}^*(i), R_{ij}^*(j))$ are the
optimum rate-power allocation between node pair $(i,j)$ given by
\cite{roumyG07jour}.
\end{itemize}
Finally, remove all nodes that do not participate in any edge. We
denote the resulting graph for a given rate allocation by
$G_M(\mathbf{R}) = (V,E,A)$, where $E$ is undirected edge set and
$A$ is directed edge set. Denote the regular node set as
$V_R\subset V$.

\begin{lemma}\label{lem:Noise GPP2path}
Consider a set of discrete correlated sources $X_1, \dots X_n$ and
a corresponding rate assignment $\mathbf{R} = (R_1, \dots, R_n)$.
Suppose that we construct $G_M(\mathbf{R})$ based on the algorithm
above. The rate assignment $\mathbf{R}$ is generalized pairwise
valid if and only if, $\forall R_i \in \mathbf{R}, Q_i(R_i) \leq
P_{max}$, and for all regular nodes $i \in V_R$, at least one of
these conditions holds:
\begin{enumerate}
\item $i$ participates in an undirected edge $(i, i^{'})$, $i'\in
V_R$; \item There exists a starred node $i^*$ and an directed edge
$(i^*\rightarrow i)$; \item There exists a starred node $j^*$ such
that there is a directed path from $j^*$ to $i$; \item There
exists a regular node $j$ participating in edge $(j, j^{'})$,
$j'\in V_R$ such that there is a directed path from $j$ to $i$;
\end{enumerate}
\end{lemma}

The proof of this lemma is very similar to that of Lemma
\ref{lem:Noiseless PP2path}. If one of the conditions 1) and 2)
holds, $X_i$ is initially decodable, and vice versa. If one of the
conditions 3) and 4) holds, $X_i$ can be decoded in a sequence of
decoding procedures which starts from an initially decodable
source $X_j$, and vice versa. Next, we introduce some definitions
crucial to the rest of the development.
\begin{definition} \label{def:head}
Given a mixed graph $G=(V,E,A)$, if $e=(i\rightarrow j)\in A$, $i$
is the tail and $j$ is the head of $e$. If $e=(i,j)\in E$, we call
both $i$ and $j$ the head of $e$. For a node $i\in V$, $h_G(i)$
denotes the number of edges for which $i$ is the head.
\end{definition}

\begin{definition}\label{def:UUG}
The \emph{underlying undirected graph} of a mixed graph $G$
denoted by  $UUG(G)$ is the undirected graph obtained from the
mixed graph by forgetting the orientations of the directed edges,
i.e., treating directed edges as undirected edges.
\end{definition}

As pointed out previously, we want some nodes to work at corner
points of two-dimensional SW region and others to work on the
slope. Thus, we need to somehow combine the two concepts of
arborescence and matching. The appropriate concept for our purpose
is the notion of a matching forest first introduced in the work of
Giles \cite{Giles1}.

\begin{definition}\label{def:mf} 
Given a mixed graph $G=(V,E,A)$, a subgraph $F$ of $G$ is called a
\emph{matching forest} \cite{Giles1} if $F$ contains no cycles in
$UUG(F)$ and any node $i\in V$ is the head of at most one edge in
$F$, i.e. $\forall i\in V, h_F(i)\leq1$.
\end{definition}

In the context of this section we also define a strict matching
forest. For a mixed graph $G$ containing regular nodes and starred
nodes, a matching forest $F$ satisfying $ h_F(i)=1, \forall i\in
V_R$ (i.e. every regular node is the head of exactly one edge) is
called a \emph{strict matching forest(SMF)}. In the noisy case,
the SMF plays a role similar to the arborescence in the noiseless
case. Now, we introduce a theorem similar to Theorem
\ref{theo:noiseless1}.

\begin{theorem}\label{theo:n1}
Given a generalized pairwise valid rate assignment $\mathbf{R}$
and corresponding power assignment $\mathbf{P}$, let
$G_M(\mathbf{R})$ be constructed as above. There exists another
valid rate assignment $\mathbf{R}^{'}$ and power assignment
$\mathbf{P}^{'}$ that can be described by the edge weights of a
strict matching forest of $G_M(\mathbf{R})$ such that
$\sum_{i=1}^n P_i^{'} \leq \sum_{i=1}^n P_i$.
\end{theorem}

\emph{Proof.} In order to find such a SMF, we first change the
weights of $G_M(\mathbf{R})$, yielding a new graph
$G_M^{'}(\mathbf{R})$. Let $W_A^{'}(i\rightarrow j), W_E^{'}(i,j)$
denote weights in $G_M^{'}(\mathbf{R})$.  Let $\Lambda$ be a
sufficiently large constant. We perform the following weight
transformation on all edges. \beq W_E^{'}(i,j) =
2\Lambda-W_E(i,j), \text{~} W_A^{'}(i\rightarrow j) =
\Lambda-W_A(i\rightarrow j). \eeq Denote the sum weight of a
subgraph $G^{'}$ of graph $G_M^{'}(\mathbf{R})$ as
$Wt_{G_M^{'}(\mathbf{R})}(G^{'})$. Next, we find a
maximum weight matching forest of $G_M^{'}(\mathbf{R})$.
which can be done in polynomial time \cite{Giles2}.
\begin{lemma}\label{lem:extSMF} The maximum weight matching forest
$F_M$ in $G_M^{'}(\mathbf{R})$ is a strict matching forest, i.e.,
it satisfies: $\forall i \in V_R, h_{F_M}(i)=1$.
\end{lemma}

\emph{Proof.} See Appendix.

Note that each regular node is head of exact one edge in $F_M$.
The power allocation is performed as follows. Any $i\in V_R$ is
the head of one of three kinds of edges in $F_M$ corresponding to
three kinds of rate-power assignment:
\begin{enumerate}
\item If $\exists (i^*\rightarrow i) \in F_M$, then set
$P_i^{'}=Q_i(H(X_i))$ and $R_i^{'}=H(X_i)$. The existence of edge
$(i^*\rightarrow i)$ in $G_M(\mathbf{R})$ means that $R_i\geq
H(X_i)$, so $R_i^{'}\leq R_i$ and $P_i^{'}\leq P_i\leq P_{max}$.
\item If $\exists (i,j) \in F_M$, set $P_i^{'}=P_{ij}^*(i)$,
$R_i^{'}=R_{ij}^*(i)$ and $P_j^{'}=P_{ij}^*(j)$,
$R_j^{'}=R_{ij}^*(j)$. The existence of edge $(i,j)$ in
$G_M(\mathbf{R})$ means that $R_i$ and $R_j$ are in the SW region,
$P_i\leq P_{max}$ and $P_j\leq P_{max}$. We know that
$P_{ij}^*(i),P_{ij}^*(j)$ is the minimum sum power solution for
node $i$ and $j$ when the rate allocation is in SW region and the
power allocation satisfies $P_{max}$ constraints. So
$P_i^{'}+P_j^{'} \leq P_i+P_j$, $P_i^{'}\leq P_{max}$,
$P_j^{'}\leq P_{max}$. \item If $\exists (j\rightarrow i)\in F_M$,
set $P_i^{'}=Q_i(H(X_i|X_j))$ and $R_i^{'}=H(X_i|X_j)$ . The
existence of edge $(j\rightarrow i)$ in $G_M(\mathbf{R})$ means
that $R_i\geq H(X_i|X_j)$, so $R_i^{'}\leq R_i$ and $P_i^{'}\leq
P_i\leq P_{max}$.
\end{enumerate}



Therefore, the new power allocation $\mathbf{P}^{'}$ reduces the
sum power. Notice that when we are assigning new rates to the
nodes, the conditions in Definition \ref{def:ged pairwise_prop}
still hold. So the new rate $\mathbf{R}^{'}$ is also valid. So
$\mathbf{P}^{'}$ is a valid power allocation with less sum power.
\endproof

The following theorem says that the valid power assignment that
minimizes the sum power can be found by finding minimum weight SMF
of an appropriately defined graph.

The graph $G^{tot}=(V^{tot},A^{tot},E^{tot})$ is such that
$V^{tot}$ consists $n$ regular nodes $1,\ldots,n$ and $n$ starred
nodes $1^*,\ldots,n^*$, and $|V^{tot}|=2n$. The directed edge set
$A^{tot}$ consists of edges $(i^*\rightarrow i),
W_A(i^*\rightarrow i)=Q_i(H(X_i))$ for $\{i:i\in[n] \hbox{ and }
Q_i(H(X_i))\leq P_{max}\}$, and directed edges $(i\rightarrow j),
W_A(i\rightarrow j)=Q_j(H(X_j|X_i))$ for $\{i,j:i,j\in[n]^2 \hbox{
and } Q_j(H(X_j|X_i))\leq P_{max}\}$. The undirected edge set
$E^{tot}$ consists of edges
$(i,j),W_E(i,j)=P_{ij}^*(i)+P_{ij}^*(j)$ for all $i,j\in[n]^2$.

Assume that $P_{max}$ is large enough so that there exist at least
one valid rate-power allocation, the following theorem shows that
the optimal rate-power allocation can be found in $G^{tot}$.
\begin{theorem}
Consider a set of sources $X_1,\ldots,X_n$. Suppose that we are
interested in finding a valid rate assignment $\mathbf{R}$ and its
corresponding power assignment $\mathbf{P}$ for these sources so
that the sum power $\sum_{i=1}^n P_i=\sum_{i=1}^n Q_i(R_i)$ is
minimum. The optimal valid power assignment can be specified by
the minimum weight SMF of $G^{tot}$.
\end{theorem}

The proof of this theorem is similar to that of Theorem
\ref{theo:noiseless 2}. Note that matching is a special case of
matching forest, and is also a special case of SMF in our problem.
Therefore, minimum weight SMF solution is always no worse than
minimum matching solution.

We now show that the minimum SMF in $G^{tot}$ can be found by
finding maximum matching forest in another mixed graph after
weight transformation. We can perform the same weight
transformation for $G^{tot}$ as we did for $G_M(\mathbf{R})$.
Denote the resulting graph as $G^{tot'}$. Find the maximum weight
matching forest $F_M^{'}$ in $G^{tot'}$. Denote the corresponding
matching forest in $G^{tot}$ as $F_M$. We claim that both
$F_M^{'}$ and $F_M$ are SMFs. To see this, note that since there
exists valid rate allocation $\mathbf{R}$, $G_M^{'}(\mathbf{R})$
is a subgraph of $G^{tot'}$. From Lemma \ref{lem:extSMF}, we know
that SMF exists in $G_M^{'}(\mathbf{R})$. Therefore, SMF also
exists in $G^{tot'}$. Because in a SMF starred node is not head of
any edge and regular node is head of exact one edge, based on
weight transformation rules, the weight of a SMF $F_S^{'}$ in
$G^{tot'}$ is: \beq\label{eq:weightSMF} Wt_{G^{tot'}}(F_S^{'}) =
n\Lambda - Wt_{G^{tot}}(F_S) \eeq where $F_S$ is the corresponding
SMF in $G^{tot}$. Weight of any non-strict matching forest
$F_{NS}$ is $Wt_{G^{tot'}}(F_{NS}^{'}) = m\Lambda -
Wt_{G^{tot}}(F_{NS}), m<n$. Since $\Lambda$ is sufficiently large,
$Wt_{G^{tot'}}(F_S^{'})>Wt_{G^{tot'}}(F_{NS}^{'})$, i.e., SMFs in
$G^{tot}$ always have larger weights. Therefore, the maximum
weight matching forest $F_M^{'}$ in $G^{tot'}$ is SMF. So is the
corresponding matching forest $F_M$ in $G^{tot}$. From
\eqref{eq:weightSMF}, it is easy to see in $G^{tot}$ the matching
forest corresponding to $F_M^{'}$ (the maximum weight matching
forest in $G^{tot'}$)  has minimum weight, i.e., $F_M$ is the
minimum SMF in $G^{tot}$.

\section{Numerical results}
\label{sec:results} We consider a wireless sensor network example
in a square area where the coordinates of the sensors are randomly
chosen and uniformly distributed in $[0,1]$. The sources are
assumed to be jointly Gaussian distributed such that each source
has zero mean and unit variance (this model was also used in
\cite{CristescuB06}). The off-diagonal elements of the covariance
matrix $\mathbf{K}$ are given by $K_{ij} = \exp (-cd_{ij})$, where
$d_{ij}$ is the distance between node $i$ and $j$, i.e., the nodes
far from each other are less correlated.  The parameter $c$
indicates the spatial correlation in the data. A lower value of
$c$ indicates higher correlation. The individual entropy of each
source is $H_1 = \frac{1}{2} \log(2\pi e \sigma^2) = 2.05$.

\begin{figure}[h] \begin{center}
  \includegraphics[width=80mm, clip=true]{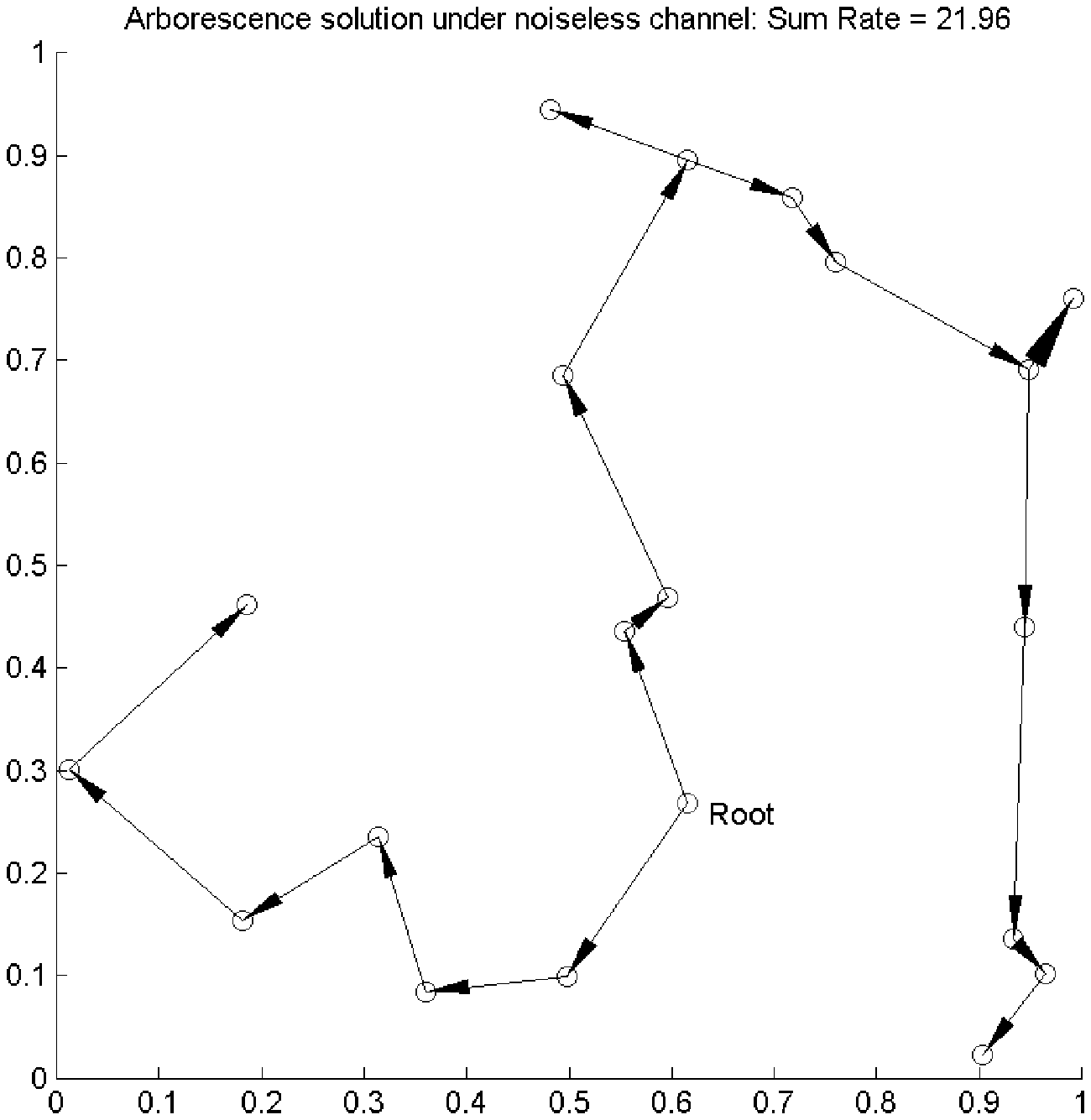}
  \caption{\label{fig:noiselessMST} Minimum arborescence solution in a WSN with 20 nodes. Noiseless channels are assumed. Correlation parameter $c=1$. Sum rate given by MA equals to 21.96, which is less than sum rate given by matching. The theoretical optimal sum rate is 20.54.}
  \end{center}
\end{figure}

\begin{figure}[h] \begin{center}
  \includegraphics[width=80mm, clip=true]{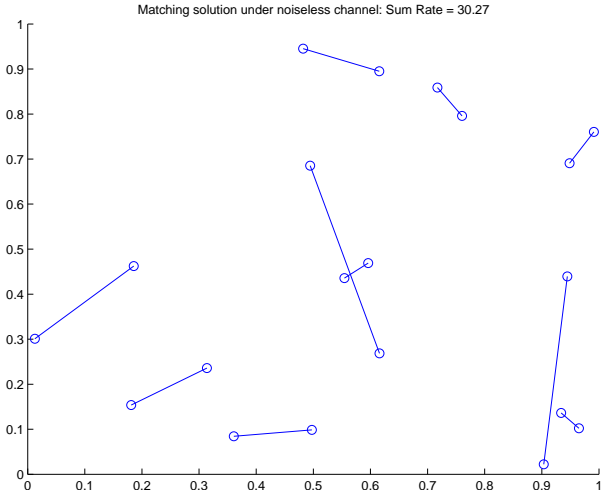}
  \caption{\label{fig:noiselessMat} Minimum matching solution in the same WSN as Fig.\ref{fig:noiselessMST}. Noiseless channels are assumed. Correlation parameter $c=1$. Sum rate given by matching equals to 30.27. Note that if we do not take advantage of correlation and transmit data individually, the sum rate will be $20\times H_1=40.94$. }
  \end{center}
\end{figure}

\begin{figure}[h] \begin{center}
  \includegraphics[width=80mm, clip=true]{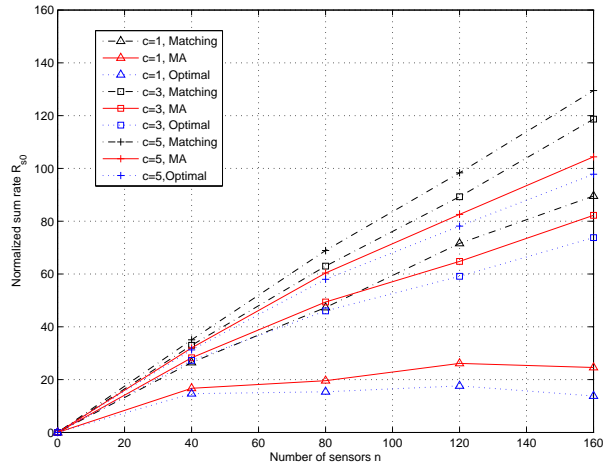}
  \caption{\label{fig:sumrate} Normalized sum rate vs. number of sensors }
  \end{center}
\end{figure}

Consider the noiseless case first. Because the rate allocation
only depends on entropies and conditional entropies, we do not
need to care the location of the sink. It is easy to see based on
our assumed model that $H(X_i|X_j)=H(X_j|X_i), \forall i,j\in
[n]^2$. Thus, $W_A(i\rightarrow j)=W_A(j\rightarrow i)$. It can be
shown that the weights of minimum weight arborescences $G_{i*},
i=1,\ldots, n$ are the same. Therefore, we only need to find
minimum weight arborescence on $G_{1^*}$.  A solution for a sensor
network containing 20 nodes are shown in
Fig.\ref{fig:noiselessMST}. Since the starred node $1^*$ is
virtual in the network, we did not put it on the graph. Instead,
we marked node 1 as root in the arborescence, whose transmission
rate is its individual entropy $H_1$. Edge $(i\rightarrow j)$ in
the arborescence implies that $X_i$ will be decoded in advance and
used as side information to help decode $X_j$.  The matching
solution for the same network is shown in
Fig.\ref{fig:noiselessMat}. As noted in \cite{roumyG07jour}, the
optimum matching tries to match close neighbors together because
$H(X_i,X_j)$ decreases with the internode distance. Our
arborescence solution also showed similar property, i.e., a node
tended to help its close neighbor since the conditional entropies
between them are small. In Fig.\ref{fig:sumrate}, we plot the
normalized sum rate $R_{s0}\triangleq \sum_{i=1}^n R_i / H_1$ vs.
the number of sensors $n$. If there is no pairwise decoding, i.e.,
the nodes transmits data individually to the sink, $R_i=H_1$ and
$R_{s0}=n$. The matching solution and the minimum arborescence
(MA) solution are compared in the figure. We also plotted the
optimal normalized sum rate $H(X_1,\ldots, H_n)/H_1$ in the
figure. The rate can be achieved theoretically when all sources
are jointly decoded together. We observe that if the nodes are
highly correlated $(c=1)$, the present solution outperforms the
matching solution considerably. Even if the correlation is not
high, our MA solution is always better than matching solution. It
is interesting to note that even though we are doing pairwise
distributed source coding, our sum rate is quite close to the
theoretical limit which is achieved by $n$-dimensional distributed
source coding.

\begin{figure}[h] \begin{center}
  \includegraphics[width=80mm, clip=true]{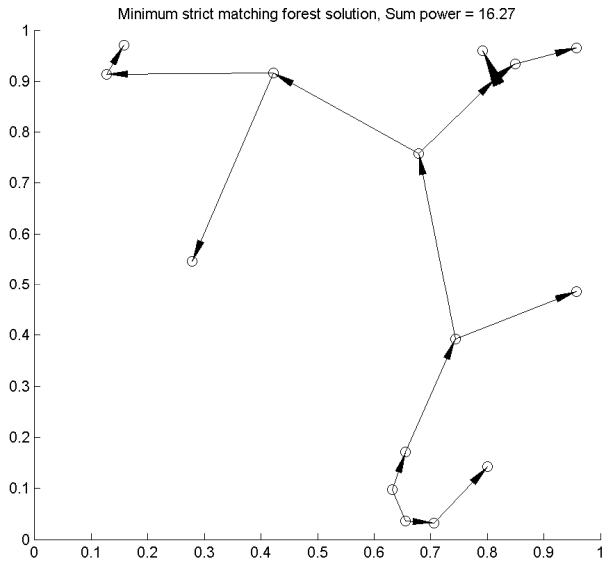}
  \caption{\label{fig:noisySMF} Minimum strict matching forest solution in a WSN with 16 nodes. AWGN channels are assumed. Correlation parameter $c=1$. Peak power constraint $P_{max}=10$.  Sum power given by SMF equals to 16.27. The optimal sum power when we apply $n$-dimensional SW codes is 14.06. }
  \end{center}
\end{figure}

\begin{figure}[h] \begin{center}
  \includegraphics[width=80mm, clip=true]{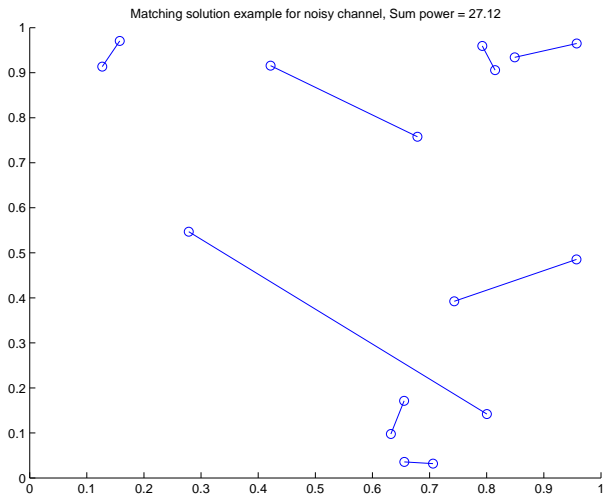}
  \caption{\label{fig:noisyMat} Minimum matching solution in the same WSN as Fig.\ref{fig:noisySMF}. AWGN channels are assumed. Correlation parameter $c=1$. Peak power constraint $P_{max}=10$. Sum power given by matching equals to 27.12. Note that if we do not take advantage of correlation and transmit data individually, the sum power will be 47.11.
            }
  \end{center}
\end{figure}

Next, we consider optimizing the total power when there are AWGN
channels between the sources and the sink. The channel gain
$\gamma_i$ is the reciprocal of the square of the distance between
source $X_i$ and the sink. We assume that the coordinates of the
sink are $(0,0)$. An example of the strict matching forest (SMF)
solution to a network with 16 sensors is given in
Fig.\ref{fig:noisySMF}. There is one undirected edge in the SMF
implying that the heads of this edge work on the slope of SW
region. Other 14 edges are directed edges implying that the tails
of the edges are used as side information to help decode their
heads. No node is encoded at rate $H_1$. In fact, most minimum
SMFs in our simulations exhibit this property, i.e., the minimum
SMF contains $1$ undirected edge and $n-2$ directed edges between
regular nodes. This fact coincides our intuition: transmitting at
a rate of conditional entropy is the most economical way, while
transmitting at a rate of individual entropy consumes most power.
The matching solution for the same network is given in
Fig.\ref{fig:noisyMat}. We compare sum powers of the SMF solution
with matching solution in Table.\ref{tab:compareNoise}. The sum
powers were averaged over three realizations of sensor networks.
We also found the theoretical optimal sum power when
$n$-dimensional distributed source coding is applied by solving
the following convex optimization problem.

\bea \min_{R_1,\ldots,R_n} \sum_{i=1}^n P_i = \sum_{i=1}^n (2^{R_i} - 1)/\gamma_i \nonumber \\
\text{subject to } (2^{R_i} - 1)/\gamma_i \leq P_{max}, \forall i\nonumber \\
(R_1,\ldots,R_n) \in SW_n \nonumber \eea where $SW_n$ is the
$n$-dimensional Slepian-Wolf region. From the table, we can
observe that our strategy always outperforms the matching strategy
regardless of the level of correlation, and comes quite close to
the theoretical limit that is achieved by $n$-dimensional SW
coding.

\section{Conclusion}
\label{sec:conclusion} The optimal rate and power allocation for a
sensor network under pairwise distributed source coding constraint
was first introduced in \cite{roumyG07jour}. We proposed a more
general definition of pairwise distributed source coding and
provided solutions for the rate and power allocation problem,
which can reduce the cost (sum rate or sum power) further. For the
case when the sources and the terminal are connected by noiseless
channels, we found a rate allocation with the minimum sum rate
given by the minimum weight arborescence on a well-defined
directed graph. For noisy orthogonal source terminal channels, we
found a rate-power allocation with minimum sum power given by the
minimum weight strict matching forest on a well-defined mixed
graph. All algorithms introduced have polynomial-time complexity.
Numerical results show that our solution has significant gains
over the solution in \cite{roumyG07jour}, especially when
correlations are high.

Future research directions would include extensions to resource
allocation problems when joint decoding of three (or more) sources
\cite{LiverisLNXG03} at one time is considered, instead of only
two in this paper. Another interesting issue is to consider
intermediate relay nodes in the network, which are able to copy
and forward data, or even encode data using network coding
\cite{al}.

\section{Acknowledgements}
The authors would like to thank the anonymous reviewers whose
comments greatly improved the quality of the paper.

\begin{appendix}

\begin{figure}[h]
\begin{center}
  \includegraphics[width=80mm, clip=true]{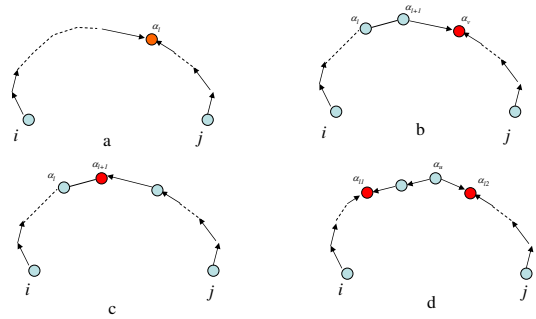}
  \caption{\label{fig:case2a} Case 2a: When $h_{F_M}(i)=0, h_{F_M}(j)=0$, path $i-\alpha_1-\alpha_2-\cdots - j$ can not exists in $UUG(F_M)$ because it will cause at lease one node $\alpha_l$, $h_{F_M}(\alpha_l)=2$.}
  \end{center}
\end{figure}

\begin{center}
PROOF OF LEMMA \ref{lem:extSMF}
\end{center}
We shall first introduce and prove a lemma which facilitates the
proof of Lemma \ref{lem:extSMF}.
\begin{lemma}\label{lem:nopathUUG}
Consider two nodes $i$ and $j$ in a matching forest $F$ such that
either $h_F(i)=0$ or $h_F(j)=0$, and they do not have incoming
directed edges. Then, there does not exist a path of the form
\beq\label{eq:UUGpathlem} i-\alpha_1-\alpha_2-\cdots-\alpha_k-j
\eeq in $UUG(F)$.
\end{lemma}
\emph{Proof.} First consider the case when $h_F(i)=h_F(j)=0$,
i.e., $i,j$ only have outgoing directed edge(s). Suppose there is
such a path \eqref{eq:UUGpathlem}, edge $(i,\alpha_1)$ should
directed from $i$ to $\alpha_1$ in $F$ since $h_{F}(i)=0$,
similarly, $j\rightarrow \alpha_k$. As depicted in
Fig.\ref{fig:case2a},  at least one node $\alpha_l$ in the path
will have $h_{F}(\alpha_l)=2$. But we know that $h_{F}(t)\leq1$
holds for every node $t\in V$ in matching forest $F$. So there is
no such path \eqref{eq:UUGpathlem} in $UUG(F)$. If
$h_F(i)=0,h_F(j)=1$ and $j$ connects to an undirected edge
$(j,j')$ in $F$, $i, j$ and $j'$ can only have outgoing directed
edge(s). By similar arguments above, we know that at least one
node $\alpha_l$ on the path is such that $h_F(\alpha_l)=2$.
Similarly, the case when $i$ connects to an undirected edge and
$h_F(j)=0$ can be proved. \endproof

\begin{figure}[h]
\begin{center}
  \includegraphics[width=80mm, clip=true]{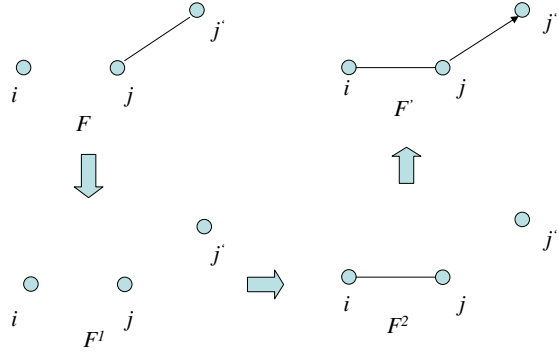}
  \caption{\label{fig:case2b2} Case 2$b_2$ When $h_{F_M}(i)=0, h_{F_M}(j)=1, (j,j^{'})\in F_M$, by introducing two intermediate matching forest $F^{1}$, $F^{2}$, we can find a new matching forest $F^{'}$ with larger sum weight.}
  \end{center}
\end{figure}

\emph{Proof of Lemma \ref{lem:extSMF}:} We will prove this lemma by contradiction. We shall show that if $h_{F_M}(i)=0$ for a regular node $i$, we can find another matching forest $F^{'}$ in  $G_M^{'}(\mathbf{R})$ such that $Wt_{G_M^{'}(\mathbf{R})}(F^{'})>Wt_{G_M^{'}(\mathbf{R})}(F_M)$, i.e., $F_M$ is not the maximum matching forest. Since $F_M$ is a matching forest, it satisfies (a) $h_{F_M}(t)\leq 1$ for every node\footnote{Actually, for a star node $i^{*}\in V\backslash V_R$, $h_{F}(i^{*})=0$ in all matching forest $F$ of $G_M^{'}(\mathbf{R})$ because there is no incoming edge to $i^{*}$ and $i^{*}$ does not participate in any undirected edge.} $t\in V$ and (b) no cycle exist in $UUG(F_M)$. Suppose $h_{F_M}(i)=0$ for a regular node $i$ in $F_M$. We shall make a set of modifications to $F_M$ resulting in a new matching forest $F^{'}$ and prove that these manipulations will eventually increase the sum weight, make $h_{F^{'}}(i)$ become 1 and ensure that there is no cycle in $UUG(F^{'})$. Also, these modifications should guarantee that $h_{F^{'}}(j)=1$ for $j\in\{j:j\in V_R\backslash \{i\} \text{ and } h_{F_M}(j)=1\}$, i.e. nodes that were previously the head of some edge continue to remain that way.
 During the proof, we shall use the properties of $G_M^{'}(\mathbf{R})$ given in Lemma \ref{lem:Noise GPP2path}. Since $\mathbf{R}$ is valid, regular node $i$ has at least one of those four properties in $G_M^{'}(\mathbf{R})$. We shall discuss these cases in a more detailed manner:

\begin{list}{}{\leftmargin=0em}

\item {\it Case 1}. If there exists a directed edge
$(i^*\rightarrow i)$ in $G_M^{'}(\mathbf{R})$, add this edge to
$F_M$ to form $F'$. Clearly,
$Wt_{G_M^{'}(\mathbf{R})}(F')>Wt_{G_M^{'}(\mathbf{R})}(F_M)$.
Since there is only one outgoing edge from $i^*$ and it has no
incoming edge, no cycle in $UUG(F^{'})$ is produced in our
procedure. And $h_{F^{'}}(t)\leq1$ still holds for every node
$t\in V$, so $F^{'}$ is still a matching forest.

\item {\it Case 2}. If there exists an undirected edge $(i,j)$ in
$G_M^{'}(\mathbf{R})$, we can include this edge to $F_M$ to
increase sum weight. Here, $h_{F_M}(i)=0$ and there are two
possibilities for $h_{F_M}(j)$, 0 or 1.
\begin{list}{}{\leftmargin=0em}
\item {\it Case 2a}. If $h_{F_M}(j)=0$, add undirected edge
$(i,j)$ to $F_M$, resulting a new subgraph $F^{'}$. Obviously, the
sum weight is increased while adding one edge. Since
$h_{F_M}(i)=h_{F_M}(j)=0$, by Lemma \ref{lem:nopathUUG} there does
not exist path with form \eqref{eq:UUGpathlem} in $UUG(F_M)$.
Thus, adding $(i,j)$ does not introduce cycle in $UUG(F')$.
$F^{'}$ is a matching forest.

\item {\it Case 2b}. If $h_{F_M}(j)=1$, we still add $(i,j)$ but
need to perform some preprocessing steps. Based on what kind of
edge connects to node $j$, we have two cases:
\begin{list}{}{\leftmargin=0em}
\item {\it Case 2$b_1$}. If there exists one directed edge $(j^{'}
\rightarrow j)$ in $F_M$, delete edge $(j^{'} \rightarrow j)$, we
have an intermediate matching forest $F^{''}$ such that
$h_F{''}(j)=0$. Add the undirected edge $(i,j)$ to obtain $F^{'}$.
Note that $F^{'}$ is a matching forest because of arguments in
Case 2a and
$Wt_{G_M^{'}(\mathbf{R})}(F^{'})>Wt_{G_M^{'}(\mathbf{R})}(F_M)$
because for a sufficient large $\Lambda$,
$2\Lambda-W_E(i,j)>\Lambda-W_A(j^{'}\rightarrow j)$.

\item {\it Case 2$b_2$}. If there exists one undirected edge
$(j^{'},j)$ in $F_M$, we notice that the existence of $(j^{'},j)$
in $G_M^{'}(\mathbf{R})$ indicates that $(R_{j^{'}},R_j)\in
SW_{j^{'}j}$, so $R_{j^{'}}\geq H(X_{j^{'}}|X_j)$ and $R_j\geq
H(X_j|X_{j^{'}})$ , which implies that there exist directed edges
$(j\rightarrow {j^{'}})$ and $({j^{'}}\rightarrow j)$ in
$G_M^{'}(\mathbf{R})$. So we can first delete edge $(j^{'},j)$ and
then add edges $(i,j)$ and $(j\rightarrow j^{'})$ to form $F^{'}$.
Adding $(j\rightarrow j^{'})$ is to make sure
$h_{F^{'}}(j^{'})=1$. These modifications are shown in
Fig.\ref{fig:case2b2}. After removing edge $(j^{'},j)$, we have an
intermediate matching forest $F^1$ such that $h_{F^1}(j)=0$ and
$h_{F^1}(j^{'})=0$. We add edge $(i,j)$ to obtain $F^2$. Because
of Lemma \ref{lem:nopathUUG}, $F^2$ is still a matching forest and
$h_{F^2}(j^{'})=0$. Then we add $(j\rightarrow j^{'})$ to obtain a
new subgraph $F^{'}$. From Lemma \ref{lem:nopathUUG}, we know that
    $(j\rightarrow j^{'})$ will not introduce cycle. Therefore, $F^{'}$ is still a matching forest. For a large enough $\Lambda$, $(2\Lambda-W_E(i,j))+(\Lambda-W_A(j\rightarrow j^{'}))> 2\Lambda-W_E(j,j^{'})$ 
    holds, so the sum weight will increase.


\end{list}

\end{list}

\item {\it Case 3}. If there exist a path from $h$ to $i$ in
$G_M^{'}(\mathbf{R})$:$h\rightarrow \gamma_1\rightarrow
\gamma_2\rightarrow\cdots\rightarrow \gamma_{k_1} \rightarrow i$,
where $h$ is a starred node or participates in an undirected edge
in $G_M^{'}(\mathbf{R})$, we use the following approach. Note that
$\gamma_1,\ldots,\gamma_{k_1}$ may participate in undirected
edges. On this path, we find the node $j$ closest to $i$ such that
$j$ participates in an undirected edge in $G_M^{'}(\mathbf{R})$ or
it is a starred node. $j$ may be the same as $h$ or be some
$\gamma_l$. We will focus on the path from $j$ to $i$, denoted by
$j\rightarrow \alpha_1\rightarrow
\alpha_2\rightarrow\cdots\rightarrow \alpha_k \rightarrow i $. The
basic idea is to add edge $\alpha_k\rightarrow i$ into $F_M$.
However, if we just simply add this edge, it may produce cycle in
underlying undirected graph. So we need more manipulations.

\begin{list}{}{\leftmargin=0em}
\item {\it Case 3a}. If $j$ is a starred node, denote $j$ as
$j^{*}$, we want to add the path
\begin{equation}\label{eq:pathji}
j^{*}\rightarrow \alpha_1\rightarrow
\alpha_2\rightarrow\cdots\rightarrow \alpha_k \rightarrow i
\end{equation}
to $F_M$. First, in $F_M$, remove all incoming directed edges to
$\alpha_l$ ($1\leq l\leq k$), then we have an intermediate
matching forest $F^1$. Note that $j^*$, $i$, and $\alpha_l$'s only
have outgoing edges, by Lemma \ref{lem:nopathUUG}, we know that
there does not exist undirected path with the form
$j^*(\text{or
}\alpha_{l_1})-\beta_1-\beta_2-\cdots-\beta_k-i(\text{or
}\alpha_{l_2})$
in $UUG(F^1)$ where $\beta$'s are nodes outside the path \eqref{eq:pathji}. Therefore, adding path \eqref{eq:pathji} into $F^1$ to form $F^{'}$ will not introduce a cycle. All nodes $\alpha_l (1\leq l\leq k)$ on the path, $h_{F^{'}}(\alpha_l)=1$. $F^{'}$ is a matching forest. Next we shall consider the weights. At some nodes, take $\alpha_l$ for example, although we deleted directed edge $(\alpha_{l^{'}}\rightarrow\alpha_l)$, where $\alpha_{l^{'}}$ is a node outside path \eqref{eq:pathji},  we add another directed edge $(\alpha_{l-1}\goes\alpha_l)$. The weight might decrease by $(\Lambda-W_A(\alpha_{l{'}}\rightarrow\alpha_l))-(\Lambda-W_A(\alpha_{l-1}\goes\alpha_l))$. 
Suppose we delete and add edges around $d$
nodes:$\alpha_{l_1},\alpha_{l_2},\ldots,\alpha_{l_d}$, the total
weight decrease is $\sum_{i=1}^d
W_A(\alpha_{l_i-1}\goes\alpha_{l_i})-W_A(\alpha_{l_i{'}}\rightarrow\alpha_{l_i})$.
It may be positive but it does not contain a $\Lambda$ term. At
the end, we will add $(\alpha_k \rightarrow i)$ without deleting
any edge coming into $i$ since $h_{F_M}(i)=0$, the weight will
increase $(\Lambda-W_A(\alpha_k\rightarrow i))$ by this operation.
If $\Lambda$ is large enough, the sum weight will finally
increase.

\item {\it Case 3b}. If $j$ participates in an undirected edge
$(j^{'},j)$ in $G_M^{'}(\mathbf{R})$. Note that $j^{'}\neq
\alpha_1,\ldots,\alpha_k$ since $j$ is the first node in the path
that participates in an undirected edge. In this case, if
$(j^{'},j)$ is already in $F_M$, we just need to add the path
\eqref{eq:pathji} from $j$ to $i$ as we did in the case above to
form $F^{'}$. The resulting path is :
    $j^{'}-j\rightarrow \alpha_1\rightarrow \alpha_2\rightarrow\cdots\rightarrow \alpha_k \rightarrow i$
    Note that in $F_M$, $j^{'},j$ do not have directed incoming edges. By similar argument in the previous case, we know that $F^{'}$ is a matching forest. If $(j^{'},j)$ is not in $F_M$, we want to add  $(j^{'},j)$ to $F_M$ and then add the path \eqref{eq:pathji}. We have four possibilities, some of which require preprocessing:

\begin{list}{}{\leftmargin=0em}
\item {\it Case $3b_1$}.  $h_{F_M}(j)=0$ and $h_{F_M}(j^{'})=0$;
we can add $(j^{'},j)$ as we did in Case 2a, and then we add path
\eqref{eq:pathji} as we did above. \item {\it Case $3b_2$}.
$h_{F_M}(j)=0$ and $h_{F_M}(j^{'})=1$; we can add $(j^{'},j)$
after some preprocessing as we did in Case 2$b_1$ and Case 2$b_2$,
and then we add path \eqref{eq:pathji} as we did above.

 Next we discuss cases in which $h_{F_M}(j)=1$. In this case, we only need to consider some directed edge $(j^{''}\rightarrow j)$ comes into $j$ in $F_M$. If there some undirected edge $(j^{''},j)$ connecting $j$ in $F_M$, this case has been discussed in Case $3b$ above, by treating $j^{''}$ as $j^{'}$.

\item {\it Case $3b_3$}. $h_{F_M}(j)=1, (j^{''}\rightarrow j)$,
and $h_{F_M}(j^{'})=0$; We can delete $(j^{''}\rightarrow j)$ and
add $(j,j^{'})$ as we did in Case 2$b_1$, node $j^{'}$ is regarded
as $i$ in Case 2$b_1$, it is guaranteed that the resulting
subgraph is a matching forest. And then we add path
\eqref{eq:pathji} as we did above.

\item {\it Case $3b_4$}. $h_{F_M}(j)=1, (j^{''}\rightarrow j)$,
and $h_{F_M}(j^{'})=1$; For $j^{'}$, it could be head of an
undirected edge or a directed edge. If $j^{'}$ is head of an
undirected edge $(j^{'},j^{'''})$, we perform operations shown in
Fig.\ref{fig:case3b41} to get $F'$. The possible weight decrease
during our operations around node $j$ is $(W_A(j^{'}\rightarrow
j^{'''})-W_A(j^{''}\rightarrow
j))+((W_E(j,j^{'})-W_E(j^{'},j^{'''}))$. We will add edge
$(\alpha_k\rightarrow i)$ on path \eqref{eq:pathji} with weight
$\Lambda-W_A(\alpha_k\rightarrow i)$. Since $\Lambda$ is large
enough, the sum weight will still increase. If $j^{'}$ is head of
a directed edge $(j^{'''}\rightarrow j^{'})$, we perform
operations shown in Fig.\ref{fig:case3b42} to get $F'$. Similarly,
because $\Lambda$ is large enough, the sum weight will increase.

\end{list}
\end{list}
\end{list}

\end{appendix}

\bibliographystyle{IEEEtran}
\bibliography{tip,RGBIB}

\begin{table}[h]
\begin{center}
\caption{\label{tab:compareNoise} Comparison of sum powers between
minimum strict matching forest and matching solution.
$P_{max}=10$.}
    \begin{tabular}{|c|c|c|c|c|}
    \hline
    \multicolumn{2}{|c|}{Number of nodes} & 4 & 8 & 12\\ \hline
    \multirow{3}{*} {$c=1$}
    & SMF       & 5.57 & 7.49 & 11.17 \\
    & Matching & 6.20 & 10.71 & 16.99 \\
    & Optimal  & 5.45 & 7.06 & 9.93 \\ \hline

    \multirow{3}{*} {$c=3$}
    & SMF       & 6.22 & 16.72 & 21.15 \\
    & Matching & 6.30 & 17.81 & 23.79 \\
    & Optimal & 6.17 & 16.44 & 20.60 \\ \hline
    \multirow{3}{*} {$c=5$}
    & SMF       & 9.68 & 18.65  & 25.14 \\
    & Matching & 9.92 & 18.91 & 25.83\\
    & Optimal & 9.67 &18.56 & 24.96 \\ \hline

    \end{tabular}
\end{center}

\end{table}

\begin{figure}[h]
\begin{center}
  \includegraphics[width=80mm, clip=true]{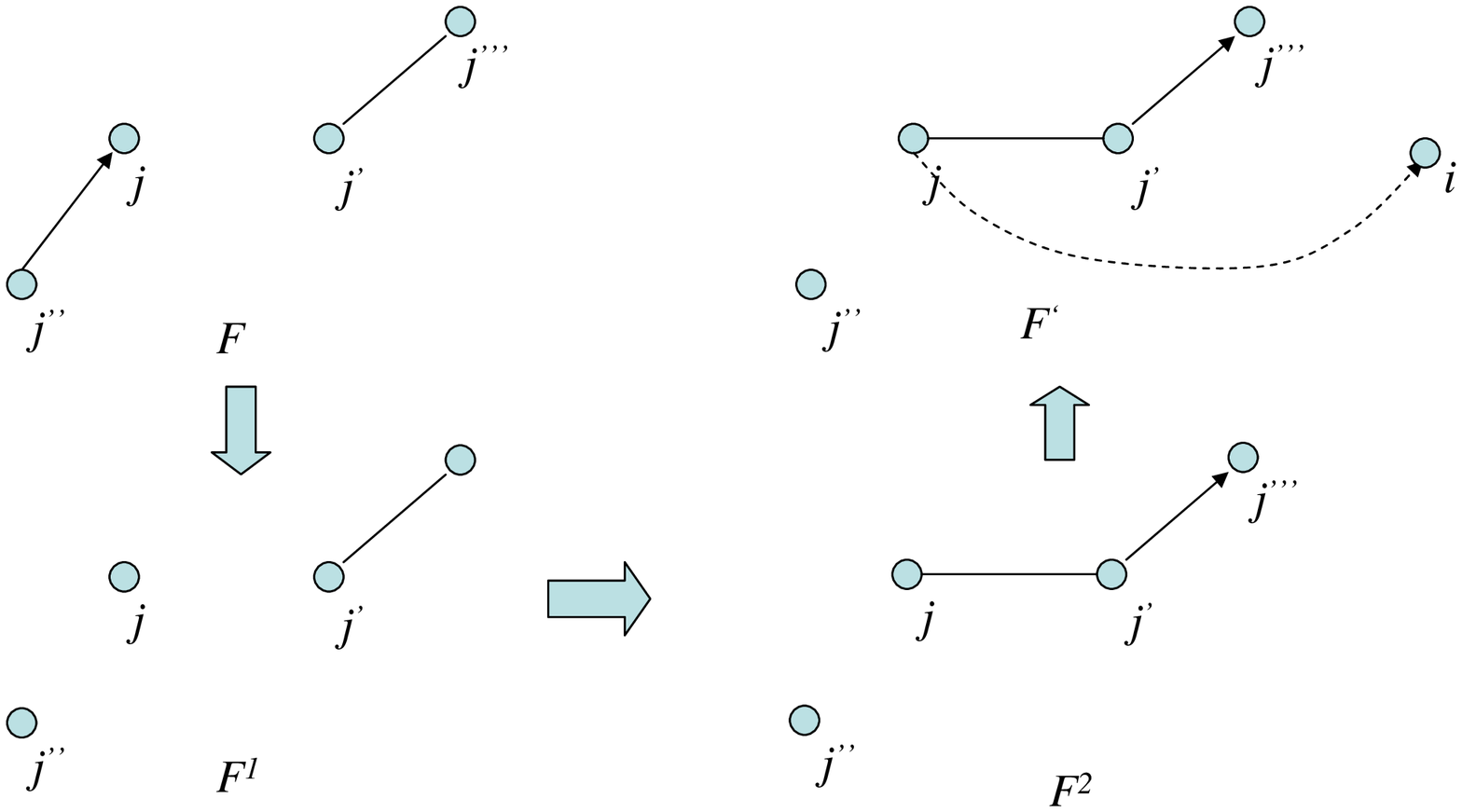}
  \caption{\label{fig:case3b41} Case $3b_{4-1}$: When $h_{F_M}(j)=h_{F_M}(j^{'})=1, (j^{'},j)\in G_M^{'}(\mathbf{R}), (j^{''}\goes j)\in F_M,(j^{'},j^{'''})\in F_M$, remove $(j^{''}\goes j)$ to form an intermediate matching forest $F^{1}$ where $h_{F^{1}}(j)=0, h_{F^{1}}(j^{'})=1, \text{ and } (j^{'},j^{'''})\in F^{1}$. Then apply the same operations as case$(2b_2)$, resulting another matching forest $F^{2}$. Finally add the path from $j$ to $i$ to get $F^{'}$.}
  \end{center}
\end{figure}

\begin{figure}[h]
\begin{center}
  \includegraphics[width=80mm, clip=true]{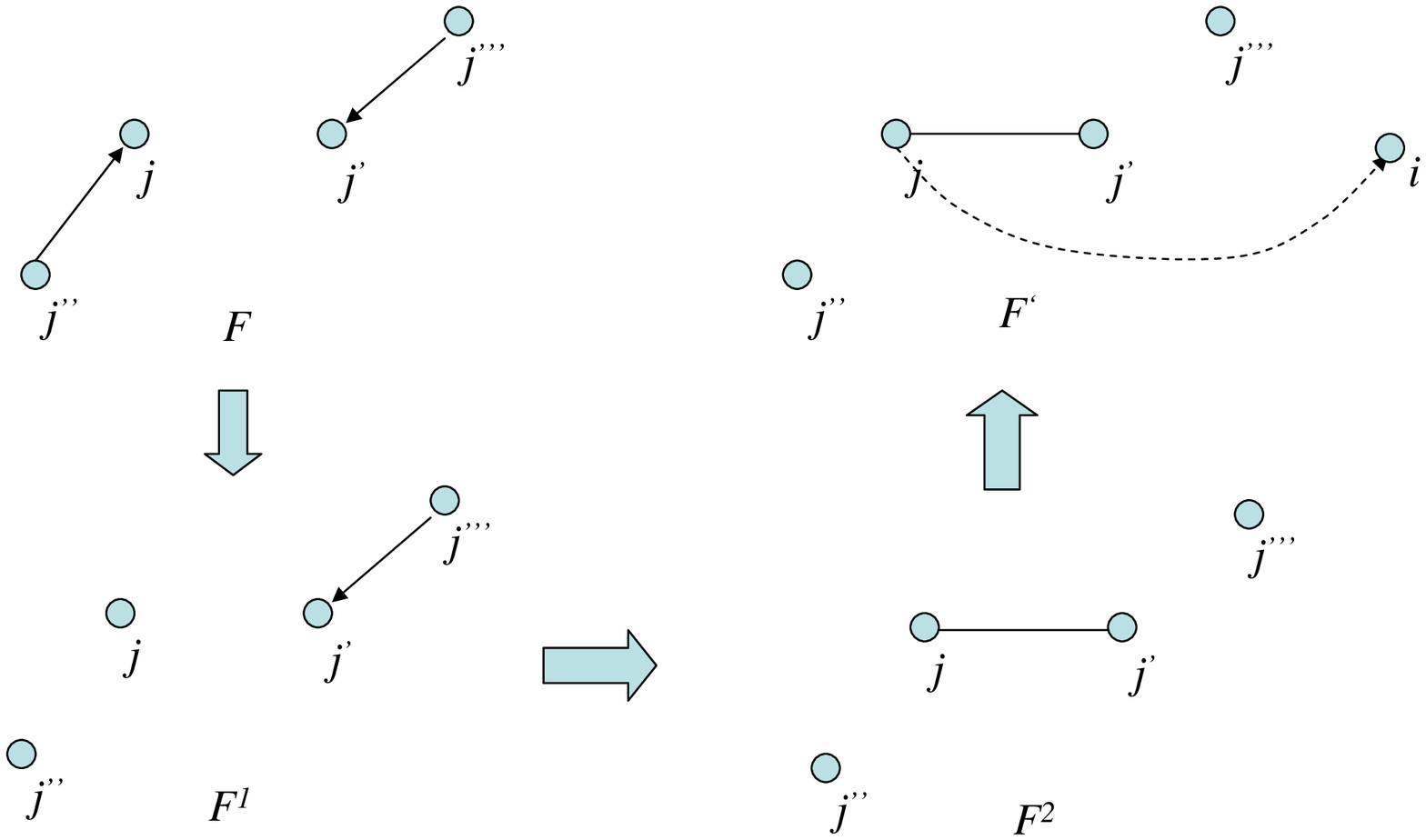}
  \caption{\label{fig:case3b42} Case $3b_{4-2}$: when $h_{F_M}(j)=h_{F_M}(j^{'})=1, (j^{'},j)\in G_M^{'}(\mathbf{R}), (j^{''}\goes j)\in F_M,(j^{'''}\goes j^{'})\in F_M$, remove $(j^{''}\goes j)$ to form an intermediate matching forest $F^{1}$ where $h_{F^{1}}(j)=0, h_{F^{1}}(j^{'})=1, \text{ and } (j^{'''}\goes j^{'})\in F^{1}$. Then apply the same operations as case$(2b_1)$, resulting another matching forest $F^{2}$. Finally add the path from $j$ to $i$ to get $F^{'}$.}
  \end{center}
\end{figure}

\end{document}